\documentstyle[12pt]{article}
\begin{document}
\begin{flushright}
Preprint IHEP 94-75\\
July 26, 1994
\end{flushright}
\begin{center}
{\bf \large Scaling Properties of $S$-wave Level Density\\for Heavy Quarkonium
from QCD Sum Rules}
\vspace*{1cm}
{V.V.Kiselev}\\
{\it Institute fir High Energy Physics,\\
Protvino, Moscow Region, 142284, Russia\\
Fax: +7-095-230-23-37}
\end{center}
\begin{abstract}
In the framework of a specific scheme of the QCD sum rules
for $S$-wave levels of the heavy quarkonium, one derives an expression,
relating the energetic density of quarkonium states and universal
characteristics in the heavy quarkonium physics, such as the difference
between the masses of a heavy quark $Q$ and meson $(Q\bar q)$ and the number
of heavy quarkonium levels below the threshold of $(Q\bar Q) \to
(Q\bar q) + (\bar Q q)$ decay.
\end{abstract}

\section*{Introduction}

Powerful tools in studies of heavy quarkonia, the bound states of two heavy
quarks, are phenomenological potential models \cite{1,2,3} and QCD sum rules
\cite{4}. An applicability of the approaches to the systems of two heavy
quarks is caused by

1) a low value of the ratio $\Lambda_{QCD}/m_Q \ll 1$, where $m_Q$ is the
heavy quark mass and $\Lambda_{QCD}$ is the quark confinement scale,
determining the inverse distance between the quarks in the bound states, and

2) a nonrelativistic motion of the heavy quarks inside the quarkonium,
$v \to 0$.

In the QCD sum rules, the low value of ratio $\Lambda_{QCD}/m_Q$ determines
not large contribution of higher orders of the QCD perturbation theory over the
quark-gluon coupling $\alpha_S \sim 1/\ln(m_Q/\Lambda_{QCD})\ll 1$ in
the expansion of Wilson's coefficients, and it makes a suppression
of nonperturbative quark-gluon condensate contribution, having the power form,
as $O(\Lambda^4_{QCD}/m_Q^2)$ for the gluonic condensate $<\alpha_S \;
G^2_{\mu\nu}>$ contribution into the sum rules for vector currents, for
example.

In the potential models, from data on the spectroscopy of the $(\bar c c)$
charmonium and the $(\bar b b)$ bottomonium one finds that the nonrelativistic
quark motion allows one to get the phenomenological potential in the range of
average distances between the heavy quarks inside the quarkonia
\begin{equation}
0.1\; fm < r < 1\;fm\;. \label{n1}
\end{equation}

Being the potential of static sources for the gluon field, this potential must
not depend on the flavours of sources. This flavour-independence is empirically
confirmed for the QCD motivated potentials \cite{1}. Such potentials,
possessing different asymptotic properties in the regions of $r \to 0$ and
$r \to \infty$, coincide with each other in the region (\ref{n1}), where
they allow approximations, having a simple scaling behaviour. These
approximations are the logarithmic \cite{2} and power \cite{3} laws
\begin{eqnarray}
V_L(r) & = & c_L + d_L \ln(\Lambda_L r)\;, \\
V_M(r) & = & c_M + d_M (\Lambda_L r)^k\;.
\end{eqnarray}
By the virial theorem
\begin{equation}
<T> = \frac{1}{2} <r \frac{dV}{dr}>\;,
\end{equation}
one finds
\begin{eqnarray}
<T_L> & = & d_L/2 = const.\;, \label{99n}\\
<T_M> & = & \frac{k}{k+2}(-c_M+E)\;, \label{10n}
\end{eqnarray}
where $E$ is the binding energy of quarks in the quarkonium.
Phenomenologically, one has $k\ll 1$, $|E|\ll |c_M|$, so that in the region
of average distances between the heavy quarks in the heavy quarkonium
(\ref{n1}), the kinetic energy of quarks practically is a constant value,
independent of the quark flavours,
\begin{equation}
<T_M> \simeq const. \label{11n}
\end{equation}
Then from the Feynman-Hellmann theorem
\begin{equation}
\frac{dE}{d\mu} = -\;\frac{<T>}{\mu}\;, \label{12n}
\end{equation}
where $\mu$ is the reduced mass of heavy quark system $(Q\bar Q')$,
one can get that the level difference in the system does not depend
on the reduced mass of quarks, i.e. on the quark flavours,
\begin{equation}
E(\bar n, \mu) - E(n,\mu) = E(\bar n, \mu') - E(n, \mu')\;. \label{13n}
\end{equation}
Condition (\ref{13n}) means that the energetic density of heavy quarkonium
levels does not depend on the quark flavours
\begin{equation}
\frac{dn}{dM_n} = \phi(n)\;,\label{14n}
\end{equation}
where $\phi(n)$ does not depend on $\mu$.

The described preprties of heavy quark potential are found phenomenologically.
They cause the high accuracy of potential models for calculations of
the heavy quarkonium masses with no account of spin-dependent splittings,
$\delta m(nL) \simeq 30$ MeV.

The accuracy for the predictions of quarkonium wave functions in the framework
of the potential models is low, for example, it is $\delta \Psi(0)/ \Psi(0)
\sim 30\div 50 \%$, since in this case the potential behaviour in the border
points ($r\to 0$ and $r\to \infty$) becomes essential.

In the QCD sum rules, the accuracy of predictions for the heavy quarkonium
masses is one order of magnitude lower than the accuracy of potential models,
$\delta m_{SR} \sim 200\div 300$ MeV. This fact is connected to that the
consideration in the QCD sum rules takes a finite number of terms in the
QCD perturbation theory for the Wilson's coefficients and a restricted set of
the quark-gluon condensates, so that the results of such noncomplete
consideration depend on an unphysical parameter, defining a scheme of
the averaging in the QCD sum rules (the number of moment for a
spectral density of current correlators or the Borel transform parameter).
An additional uncertainty is related with a modelling of a nonresonant
contribution into the current correlator, i.e. with the threshold of
hadronic continuum. Such parametric dependences lead to the low accuracy
of QCD sum rule predictions for the heavy quarkonium masses\footnote{
The QCD sum rule accuracy in calculations of the leptonic constants
($f_\psi$, $f_\Upsilon$) is higher ($\sim 20\div 25\%$), since one uses
the heavy quarkonium masses, known experimentally.}.
Moreover, the use of weight functions, defining the averaging scheme
and rapidly dropping with the energy rise, causes a suppression of the
contribution of higher excitations in the quarkonium, so that, as a result,
this contribution is neglected.

Recently the QCD sum rule scheme has been offered in papers of refs.
\cite{5,6,7,8}, so this scheme allows one to take into account the
spectroscopic characteristics of higher $S$-wave excitations. In these papers
the following universal regularities have been derived.

1) The scaling relation for the leptonic constants of $S$-wave levels of the
heavy quarkonium with the mass $M$ and the reduced quark mass $\mu$ \cite{6} is
\begin{equation}
\frac{f^2}{M}\;\biggl(\frac{M}{4\mu}\biggr)^2 = const.\;, \label{2}
\end{equation}
that is in a good agreement ($\Delta f/f \sim 5\%$) with the experimental data
\cite{9}
on the leptonic constants of $\Upsilon$-, $\psi$- and $\phi$-particles, which
are the quarkonia with the hidden flavours $(Q\bar Q)$, so that $4\mu/M=1$,
and one has \cite{5}
\begin{equation}
\frac{f^2}{M} = const.\;, \label{n2n}
\end{equation}
independently of the heavy quark flavours in the $(Q\bar Q)$ system.
Relation (\ref{n2n}) essentially differs from the scaling law for the
leptonic constants of heavy mesons $(Q\bar q)$, containing a single
heavy quark, where in Heavy Quark Effectice Theory (HQET) \cite{10}
one has
\begin{equation}
{f^2}\cdot {M} = const. \label{n3n}
\end{equation}
Law (\ref{n3n}) can be obtained from eq.(\ref{2}) in the limit
$\mu=m_q m_Q/(m_q+m_Q) \to m_q$, $m_Q \gg m_q$, $M\to m_Q$, so that
$\mu$ does not depend on the heavy quark flavour.
Eq.(\ref{2}) gives reasonable estimates for the leptonic constants of
$B$- and $D$-mesons \cite{4}, if one supposes $\mu\simeq 330$ MeV \cite{6}.

2) The scaling relation for the leptonic constants of $nS$-levels in the
quarkonium \cite{7} is
\begin{equation}
\frac{f^2_{n_1}}{f^2_{n_2}} = \frac{n_2}{n_1}\;, \label{3}
\end{equation}
independently of the heavy quark flavours. Eq.(\ref{3}) is in a good
agreement with the experimental data on the leptonic constants in the families
of $\psi$- and $\Upsilon$-particles \cite{9} ($\Delta f/f \le 10\%$).

3) The relation for the mass differences of $nS$-wave levels in the
heavy quarkonium \cite{8} is
\begin{equation}
\frac{M_n-M_1}{M_2-M_1} = \frac{\ln {n}}{\ln {2}}\;,\;\;\;n\ge 2\;,\label{m3}
\end{equation}
independently of the flavours of heavy quarks in the quarkonium. Eq.(\ref{m3})
is in a good agreement with the experimental data on the masses of
$\psi$- and $\Upsilon$-particles \cite{9} ($\delta (\Delta M)/\Delta M
\le 10\%$), too.

Regularities (\ref{2}), (\ref{3}), (\ref{m3}) allow one to improve the accuracy
of QCD sum rule results by one order of magnitude, approximately.
This improvement takes place also for the heavy quarkonium masses, where
the sum rule accuracy becomes comparable with the accuracy of potential
models, so, in addition, the sum rules allow one to get the explicit relations.

However, having derived relations (\ref{2}), (\ref{3}), (\ref{m3}), one has
used the phenomenological condition, stating the flavour-independence
of the heavy quarkonium level density and coming from the analysis, made
in the framework of the potential models.

In the present paper, in the framework of the offered scheme of QCD sum rules,
we derive the relation for the $S$-wave level density for the heavy quarkonium
\begin{equation}
\frac{dM_n}{dn}(n=1) = \frac{2\bar \Lambda}{\ln{n_{th}}}\;, \label{n16}
\end{equation}
where $\bar \Lambda = m_{(Q\bar q)} - m_Q$ is the difference between the masses
of heavy meson and heavy quark, $n_{th}$ is the number of $S$-wave levels
of the $(Q\bar Q)$ heavy quarkonium below the threshold of quarkonium decay
into the heavy meson pair $(Q\bar Q) \to (Q\bar q) + (\bar Q q)$. In the
leading order, one has
\begin{equation}
\bar \Lambda = const.\;, \label{n17}
\end{equation}
with the accuracy up to power corrections over the inverse mass of heavy quark
\cite{10} (about the role of logarithmic and power corrections, see
ref.\cite{11}).

In the leading approximation, stepping from the charmonium to the bottomonium,
one can neglect a weak logarithmic variation of the number of
levels below the thershold,
\begin{equation}
\ln{n_{th}}(b\bar b) \simeq \ln{n_{th}}(c\bar c) \;. \label{n18}
\end{equation}
{}From eq.(\ref{n16})-(\ref{n18}) it follows that in the QCD sum rules one can
show that the $S$-wave level density of heavy quarkonium does not depend
on the heavy quark flavours.

Thus, the offered scheme of QCD sum rules allows one to do the quite complete
consideration of heavy quarkonium and to use no external assumptions,
extracted from the phenomenological potential models, for example.

In Section 1 we consider the scheme of QCD sum rules with the account of
spectroscopic quantities for the heavy quarkonium and derive relation
(\ref{n16}). In Section 2 we make the phenomenological analysis of
eq.(\ref{n16}). In the Conclusion the obtained results are summarized.

\section{Heavy Quarkonium Sum Rules}

Let us consider the two-point correlator functions of quark currents
\begin{eqnarray}
\Pi_{\mu\nu} (q^2) & = & i \int d^4x e^{iqx} <0|T J_{\mu}(x)
J^{\dagger}_{\nu}(0)|0>\;,
\label{1} \\
\Pi_P (q^2) & = & i \int d^4x e^{iqx} <0|T J_5(x) J^{\dagger}_5(0)|0>\;,
\end{eqnarray}
where
\begin{eqnarray}
J_{\mu}(x) & = & \bar Q_1(x) \gamma_{\mu} Q_2(x)\;,\\
J_5(x) & = & \bar Q_1(x) \gamma_5 Q_2(x)\;,\\
\end{eqnarray}
$Q_i$ is the spinor field of the heavy quark with $i = c, b$.

Further, write down
\begin{equation}
\Pi_{\mu\nu} = \biggl(-g_{\mu\nu}+\frac{q_{\mu} q_{\nu}}{q^2}\biggr) \Pi_V(q^2)
+ \frac{q_{\mu} q_{\nu}}{q^2} \Pi_S(q^2)\;,
\end{equation}
where $\Pi_V$ and $\Pi_S$ are the vector and scalar correlator functions,
respectively. In what follows we will consider the vector and pseudoscalar
correlators: $\Pi_V(q^2)$ and $\Pi_P(q^2)$.

Define the leptonic constants $f_V$ and $f_P$
\begin{eqnarray}
<0|J_{\mu}(x) |V(\lambda)> & = & i \epsilon^{(\lambda)}_{\mu}\;
f_V M_V\;e^{ikx}\;,\\
<0|J_{5\mu}(x)|P> & = & i k_{\mu}\;f_P e^{ikx}\;,
\end{eqnarray}
where
\begin{equation}
J_{5\mu}(x)  =  \bar Q_1(x) \gamma_5 \gamma_{\mu} Q_2(x)\;,
\end{equation}
so that
\begin{equation}
<0|J_{5}(x)|P>  =  i\;\frac{f_P M_P^2}{m_1+m_2}\;e^{ikx}\;, \label{9}
\end{equation}
where $|V>$ and  $|P>$ are the state vectors of $1^-$ and $0^-$
quarkonia, and $\lambda$ is the vector quarkonium polarization, $k$
is 4-momentum of the meson, $k_{P,V}^2 = M_{P,V}^2$.

Considering the charmonium ($\psi$, $\psi '$ ...) and bottomonium ($\Upsilon$,
$\Upsilon '$, $\Upsilon ''$ ...), one can easily show that the relation
between the width of
leptonic decay $V \to e^+ e^-$  and $f_V$ has the form
\begin{equation}
\Gamma (V \to e^+ e^-) = \frac{4 \pi}{9}\;e_i^2 \alpha_{em}^2\;
\frac{f_V^2}{M_V}\;,
\end{equation}
where $e_i$ is the electric charge of quark $i$.

In the region of narrow nonoverlapping resonances, it follows from
eqs.(\ref{1}) - (\ref{9}) that
\begin{eqnarray}
\frac{1}{\pi} \Im m \Pi_V^{(res)} (q^2) & = &
\sum_n f_{Vn}^2 M_{Vn}^2 \delta(q^2-M_{Vn}^2)\;,
\label{11} \\
\frac{1}{\pi} \Im m \Pi_P^{(res)} (q^2) & = &
\sum_n f_{Pn}^2 M_{Pn}^4\;\frac{1}{(m_1+m_2)^2} \delta(q^2-M_{Pn}^2)\;.
\end{eqnarray}
Thus, for the observed spectral function one has
\begin{equation}
\frac{1}{\pi} \Im m \Pi_{V,P}^{(had)} (q^2)  = \frac{1}{\pi} \Im m
\Pi_{V,P}^{(res)} (q^2)+ \rho_{V,P}(q^2, \mu_{V,P}^2)\;,
\label{13}
\end{equation}
where $\rho (q^2,\;\mu^2)$ is the continuum contribution, which is
not equal to zero at $q^2 > \mu^2$.

Moreover, the operator product expansion gives
\begin{equation}
\Pi^{(QCD)} (q^2)  = \Pi^{(pert)} (q^2)+ C_G(q^2) <\frac{\alpha_S}{\pi} G^2> +
C_i(q^2)<m_i \bar Q_i Q_i>+ \dots\;,
\label{14}
\end{equation}
where the perturbative contribution $\Pi^{(pert)}(q^2)$ is labeled, and
the nonperturbative one is expressed in the form of sum
of quark-gluon condensates
with the Wilson's coefficients, which can be calculated in the QCD
perturbative theory.

In eq.(\ref{14}) we have been restricted by the contribution of vacuum
expectation values for the operators with dimension $d =4$.
For $C^{(P)}_G (q^2)$ one has, for instance, \cite{4}
\begin{equation}
C_G^{(P)} = \frac{1}{192 m_1 m_2}\;\frac{q^2}{\bar q^2}\;
\biggl(\frac{3(3v^2+1)(1-v^2)^2}
{2v^5} \ln \frac{1+v}{1-v} - \frac{9v^4+4v^2+3}{v^4}\biggr)\;, \label{15}
\end{equation}
where
\begin{equation}
\bar q^2 = q^2 - (m_1-m_2)^2\;,\;\;\;\;v^2 = 1-\frac{4m_1 m_2}{\bar q^2}\;.
\label{16}
\end{equation}
The analogous formulae for other Wilson's coefficients can be found in
Ref.\cite{4}. In what follows it will be clear that the explicit form
of coefficients has no significant meaning for the present consideration.

In the leading order of QCD perturbation theory it has been found for
the imaginary part of correlator that \cite{4}
\begin{eqnarray}
\Im m \Pi_V^{(pert)} (q^2) & = & \frac{\tilde s}{8 \pi s^2}
(3 \bar s s - \bar s^2 + 6m_1 m_2 s - 2 m_2^2 s) \theta(s-(m_1+m_2)^2),\\
\Im m \Pi_P^{(pert)} (q^2) & = & \frac{3 \tilde s}{8 \pi s^2}
(s - (m_1-m_2)^2) \theta(s-(m_1+m_2)^2)\;,
\end{eqnarray}
where $\bar s = s-m_1^2+m_2^2$, $ \tilde s^2 = \bar s^2 -4 m_2^2 s$.

The one-loop contribution into $\Im m \Pi(q^2)$ can be included into the
consideration (see, for example, Ref.\cite{4}). However, we note that the
more essential correction is that of summing a set over the powers of
$(\alpha_s/v)$, where $v$ is defined in eq.(\ref{16}) and is a relative quark
velocity, and $\alpha_S$ is the QCD interaction constant. In Ref.\cite{4}
it has been shown that account of the Coulomb-like gluonic
interaction between the quarks leads to the factor
\begin{equation}
F(v) = \frac{4 \pi}{3}\;\frac{\alpha_S}{v}\; \frac{1}{1-\exp (-\frac{4 \pi
\alpha_S}{3 v})}\;,
\end{equation}
so that the expansion of the $F(v)$ over $\alpha_S/v \ll 1$ restores,
precisely,
the one-loop $O(\frac{\alpha_S}{v})$ correction
\begin{equation}
F(v) \approx 1 - \frac{2 \pi}{3}\;\frac{\alpha_s}{v}\; \dots \label{20}
\end{equation}
In accordance with the dispersion relation one has the QCD sum rules,
which state that, in average, it is true that, at least, at $q^2 < 0$
\begin{equation}
\frac{1}{\pi}\;\int\frac{\Im m \Pi^{(had)}(s)}{s-q^2} ds = \Pi^{(QCD)}(q^2)\;,
 \label{21}
\end{equation}
where the necessary subtractions are omitted. $\Im m \Pi^{(had)}(q^2)$ and
 $\Pi^{(QCD)}(q^2)$ are defined by eqs.(\ref{11}) - (\ref{13}) and
eqs.(\ref{14}) - (\ref{20}), respectively.
eq.(\ref{21}) is the base to develop the sum rule approach in the forms
of the correlator function moments and of the Borel transform analysis
(see Ref.\cite{4}). The truncation of the set in the right hand side of
eq.(\ref{21}) leads to the mentioned unphysical dependence of the $f_{P,V}$
values on the external parameter of the sum rule scheme.

Further, let us use the conditions, simplifying the consideration due to
the heavy quarkonium.

\subsection{Nonperturbative Contribution}

We assume that, in the limit of the very heavy quark mass, the power
corrections of nonperturbative contribution are small. From eq.(\ref{15})
one can see that, for example,
\begin{equation}
C_G^{(P)}(q^2) \approx O(\frac{1}{m_1 m_2})\;,\;\; \Lambda/m_{1,2}\ll 1\;,
\end{equation}
where $v$ is fixed,  $q^2 \sim (m_1 + m_2)^2$,
when $\Im m \Pi^{(pert)}(q^2) \sim (m_1+m_2)^2$.
It is evident that, due to the purely dimensional consideration, one can
believe that the Wilson's coefficients tend to zero as
$1/m_{1,2}^2$.

Thus, the limit of very large heavy quark mass implies that one can neglect
the quark-gluon condensate contribution.

\subsection{Nonrelativistic Quark Motion}

The nonrelativistic quark motion implies that, in the resonant region, one has,
in accordance with eq.(\ref{16}),
\begin{equation}
v \to 0\;.
\end{equation}
So, one can easily find that in the leading order
\begin{equation}
\Im m \Pi_P^{(pert)}(s) \approx  \Im m \Pi_V^{(pert)}(s) \to \frac {3 v}
{8 \pi^2} s\; \biggl(\frac{4\mu}{M}\biggr)^2\;,
\end{equation}
so that with account of the Coulomb factor
\begin{equation}
F(v) \simeq \frac{4 \pi}{3}\; \frac{\alpha_S}{v}\;,
\end{equation}
one obtaines
\begin{equation}
\Im m \Pi_{P,V}^{(pert)}(s) \simeq \frac{\alpha_S}{2} s\;
\biggl(\frac{4\mu}{M}\biggr)^2\;. \label{27}
\end{equation}

\subsection{"Smooth Average Value" Scheme of the Sum Rules}

As for the hadronic part of the correlator, one can write down for the narrow
resonance contribution
\begin{eqnarray}
\Pi_V^{(res)}(q^2) & = & \int \frac{ds}{s-q^2}\;\sum_n f^2_{Vn} M^2_{Vn}
\delta(s-M_{Vn}^2)\;,
\label{28} \\
\Pi_P^{(res)}(q^2) & = & \int \frac{ds}{s-q^2}\;\sum_n f^2_{Pn}
\frac{M^4_{Pn}}{(m_1+m_2)^2} \delta(s-M_{Pn}^2)\;,\label{29}
\end{eqnarray}
The integrals in eqs.(\ref{28})-(\ref{29}) are simply calculated, and
this procedure is generally used.

In the presented scheme, let us introduce the function of state number
$n(s)$, so that
\begin{equation}
n(m_k^2) = k\;.
\end{equation}
This definition seems to be reasonable in the resonant region.
Then one has, for example, that
\begin{equation}
\frac{1}{\pi}\; \Im m \Pi_V^{(res)}(s) = s f^2_{Vn(s)}\; \frac{d}{ds} \sum_k
\theta(s-M^2_{Vk})\;.
\end{equation}
Further, it is evident that
\begin{equation}
\frac{d}{ds} \sum_k \theta(s-M_k^2) = \frac{dn(s)}{ds}\;\frac{d}{dn} \sum_k
\theta(n-k)\;,
\end{equation}
and eq.(\ref{28}) can be rewritten as
\begin{equation}
\Pi_V^{(res)}(q^2) = \int \frac{ds}{s-q^2}\; s f^2_{Vn(s)}\;\frac{dn(s)}{ds}\;
\frac{d}{dn} \sum_k \theta(n-k)\;.
\end{equation}
The "smooth average value" scheme means that
\begin{equation}
\Pi_V^{(res)}(q^2) = <\frac{d}{dn} \sum_k \theta(n-k)>\; \int \frac{ds}{s-q^2}
s f^2_{Vn(s)} \frac{dn(s)}{ds}\;.
\end{equation}
It is evident that, in average, the first derivative of step-like function
in the resonant region is equal to
\begin{equation}
<\frac{d}{dn} \sum_k \theta(n-k)> \simeq 1\;.
\end{equation}
Thus, in the scheme one has
\begin{eqnarray}
<\Pi_V^{(res)}(q^2)> & \approx & \int \frac{ds}{s-q^2}
s f^2_{Vn(s)}\; \frac{dn(s)}{ds}\;,
\label{35} \\
<\Pi_P^{(res)}(q^2)> & \approx & \int \frac{ds}{s-q^2}
\frac{s^2 f^2_{Pn(s)}}{(m_1+m_2)^2}\; \frac{dn(s)}{ds}\;.
\label{36}
\end{eqnarray}
Eqs.(\ref{35})-(\ref{36}) give the average correlators for the vector and
pseudoscalar mesons, therefore, due to eq.(\ref{21}) we state  that
\begin{equation}
\Im m <\Pi^{(hadr)}(q^2)> = \Im m \Pi^{(QCD)}(q^2)\;,
\end{equation}
that gives with account of eqs.(\ref{27}), (\ref{35}) and
(\ref{36}) at the physical points $s_n =M_n^2$
\begin{equation}
\frac{f_n^2}{M_n} = \frac{\alpha_S}{\pi} \; \frac{dM_n}{dn}
\; \biggl(\frac{4\mu}{M}\biggr)^2\;, \label{38}
\end{equation}
where in the limit of heavy quarks we use, that for the resonances
one has
\begin{equation}
m_1 +m_2 \approx M\;,\label{39}
\end{equation}
so that
\begin{equation}
f_{Vn} \simeq f_{Pn} = f_n\;.\label{40}
\end{equation}
Thus, one can conclude that for the heavy quarkonia the QCD sum rules give
the identity of $f_P$ and $f_V$ values for the pseudoscalar
and vector states.

Eq.(\ref{38}) differs from the ordinary sum rule scheme because it does not
contain the parameters, which are external to QCD. The quantity
$dM_n/dn$ is purely phenomenological. It defines the average mass difference
between the nearest levels with the identical quantum numbers.

Further, as it has been shown in ref.\cite{11a}, in the region of average
distances between the heavy quarks in the charmonium and the bottomonium,
\begin{equation}
0.1\; fm < r < 1\;fm\;, \label{2.1}
\end{equation}
the QCD-motivated potentials allow the approximation in the form of
logarithmic law \cite{2} with the simple scaling properties, so
\begin{equation}
\frac{dn}{dM_n} = const.\;,\label{2.2}
\end{equation}
i.e. the density of heavy quarkonium states with the given quantum
numbers do not depend on the heavy quark flavours.

In ref.\cite{5} it has been shown, that relation (\ref{2.2}) is also
practically valid for the heavy quark potential approximation by the
power law (Martin potential) \cite{3}, where, neglecting a low value of
the binding energy for the quarks inside the quarkonium, one can again
get eq.(\ref{2.2}).

In ref.\cite{5} it has been found, that relation (\ref{2.2}) is valid
 with the accuracy up to small logarithmic corrections over the
reduced mass of quarkonium, if one makes the quantization of
$S$-wave states for the quarkonium with the Martin potential by the
Bohr-Sommerfeld procedure.

Moreover, with the accuracy up to the logarithmic corrections, $\alpha_S$
is the constant value. Thus, as it has been shown in refs.\cite{5,6},
for the leptonic constants of $S$-wave quarkonia, the scaling relation
takes place
\begin{equation}
\frac{f^2}{M}\; \biggl(\frac{M}{4\mu}\biggr)^2 = const.\;, \label{2.3}
\end{equation}
independently of the heavy quark flavours.

Taking into the account eqs.(\ref{39}) and (\ref{40}) and integrating
eqs.(\ref{35}), (\ref{36}) by parts, one can get with the accuracy up to
border terms, that one has
\begin{equation}
-2f_n\; \frac{df_n}{dn}\; \frac{dn}{dM_n}\; n = \frac{\alpha_s}{\pi}\;
M_n\; \biggl(\frac{4\mu}{M_n}\biggr)^2\;. \label{2.4}
\end{equation}
Comparing eqs.(\ref{38}) and (\ref{2.4}), one finds
\begin{equation}
\frac{df_n}{f_n dn} = - \frac{1}{2n}\;, \label{2.5}
\end{equation}
that  gives, after the integration,
\begin{equation}
\frac{f^2_{n_1}}{f^2_{n_2}} = \frac{n_2}{n_1}\;. \label{2.6}
\end{equation}
Relation (\ref{2.6}) leads to that the border terms, which have been neglected
in the writing of eq.(\ref{2.4}), are identically equal to zero.

First, note that eq.(\ref{2.3}), relating the leptonic constants of
different quarkonia, turns out to be certainly valid
for the quarkonia with the hidden
flavour ($c\bar c$, $b\bar b$), where $4\mu/M=1$ \cite{5,6} (see
Table \ref{tm1}).

\begin{table}[t]
\caption{The experimental values of leptonic constants (in MeV)
for the quarkonia in comparison with the estimates of present model.}
\label{tm1}
\begin{center}
\begin{tabular}{||l|c|c||}
\hline
quantity & exp. & present \\
\hline
$f_\phi$ & $232\pm5$ & $230\pm25$ \\
$f_\psi$ & $409\pm13$ & $400\pm40$ \\
$f_\Upsilon$ & $714\pm14$ & $700\pm70$ \\
\hline
\end{tabular}
\end{center}
\end{table}

Second, eq.(\ref{2.3}) gives estimates of the leptonic constants for
the heavy $B$ and $D$ mesons, so these estimates are in a good agreement with
the values, obtained in the framework of other schemes of the QCD
sum rules \cite{4}.

Third, taking a value of the $1S$-level leptonic constant as the input one, we
have calculated the leptonic constants of higher $nS$-excitations in the
charmonium and the bottomonium and found a good agreement with the experimental
values \cite{7} (see Tables \ref{th1}, \ref{th2} and Figures \ref{fh1},
\ref{fh2}).
\begin{table}[t]
\caption{The experimental values of leptonic constants (in MeV)
for the $nS$-bottomonia in comparison with the
estimates of present model.}
\label{th1}
\begin{center}
\begin{tabular}{||c|c|c||}
\hline
quantity & exp. & present \\
\hline
$f_1$ & $715\pm15$ & input \\
$f_2$ & $487\pm16$ & $506\pm10$ \\
$f_3$ & $429\pm14$ & $412\pm8$ \\
$f_4$ & $320\pm30$ & $358\pm7$ \\
$f_5$ & $369\pm46$ & $320\pm7$ \\
$f_6$ & $240\pm30$ & $292\pm6$ \\
\hline
\end{tabular}
\end{center}
\end{table}
\begin{table}[b]
\caption{The experimental values of leptonic constants (in MeV)
for the $nS$-charmonia in comparison with the
estimates of present model.}
\label{th2}
\begin{center}
%
\end{center}
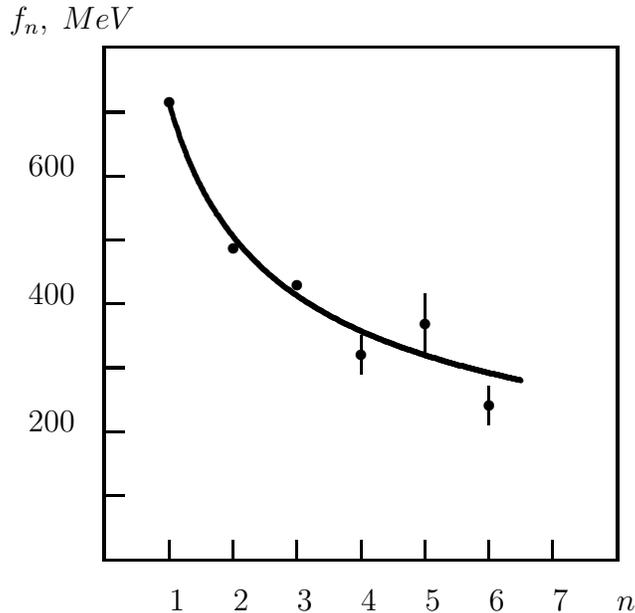
\caption{The calculated dependence of $nS$ - bottomonium leptonic
constants and the experimental values of $f_{\Upsilon(nS)}$.}
\label{fh1}
\end{figure}
\begin{figure}[t]
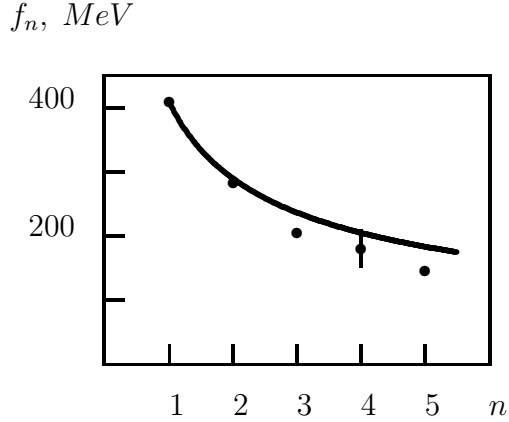

\begin{center}
%
\end{center}
\caption{The calculated dependence of $nS$ - charmonium leptonic
constants and the experimental values of $f_{\psi(nS)}$.}
\label{fh2}
\end{figure}

Further, from eqs.(\ref{38}) and (\ref{2.6}) it follows that
\begin{equation}
\frac{f_1^2}{n}\; \frac{1}{M_n} = \frac{\alpha_S}{\pi}\;
\biggl(\frac{4\mu}{M_n}\biggr)^2\; \frac{dM_n}{dn}\;, \label{2.7}
\end{equation}
so that, neglecting the low value of quark binding energy
($M_n=M_1(1+O(1/M))$), one gets
\begin{equation}
\frac{dM_n}{dn}= \frac{1}{n}\; \frac{dM_n}{dn}(n=1)\;. \label{2.8}
\end{equation}
Integrating eq.(\ref{2.8}), one partially finds eq.(\ref{m3})
\begin{equation}
\frac{M_n-M_1}{M_2-M_1} = \frac{\ln {n}}{\ln {2}}\;,\;\;\;n\ge 2\;,\label{2.9}
\end{equation}
and
\begin{equation}
{M_2-M_1} = \frac{dM_n}{dn}(n=1)\; {\ln {2}}\;.\label{2.10}
\end{equation}

Thus, in the offered scheme of QCD sum rules, one takes into account the
Coulomb-like $\alpha_S/v$-corrections and, neglecing the power corrections over
the inverse heavy quark mass, one gets the universal relation for the
differences of $nS$-wave level masses of the heavy quarkonium.

Eq.(\ref{2.9}) for the differences of $nS$-wave level masses of the heavy
quarkonium does not contain external parameters and it allows direct
comparison with the experimental data on the masses of particles in
the $\psi$- and $\Upsilon$-families \cite{9}.

Dependence (\ref{2.9}) and the experimental values for the relations of
heavy quarkonium masses are presented on Figure \ref{fm},
where one neglects the spin-spin splittings.

Note, the $\psi(3770)$ and $\psi(4040)$ charmonium states suppose to be the
results of the $3D$- and $3S$-states mixing, so that the $D$-wave dominates
in the $\psi(3770)$-state, and the mixing of the $3D$ and $3S$ wave functions
is accompanied by a small shifts of the masses, so that we have
supposed $M_3= M_{\psi(4040)}$.

As one can see from the Figure \ref{fm}, relation (\ref{2.9}) is in a
good agreement with the experimental data.
\setlength{\unitlength}{0.85mm}\thicklines
\begin{figure}[t]
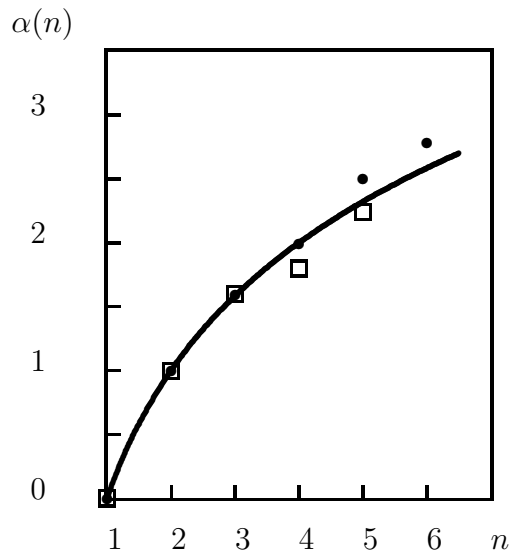

\begin{center}

\end{center}
\caption{The experimental values of $nS$ - bottomonium (solid dots)
and charmonium (empty boxes) mass differences
$\alpha(n) =(M_n-M_1)/(M_2-M_1)$
and the dependence in the present model $\alpha(n)=\ln{n}/\ln{2}$.}
\label{fm}
\end{figure}

These facts show that the offered scheme can be quite reliably applied
to the systems with the heavy quarks.

Further, using eq.(\ref{2.6}), at $q^2=0$ one can write down
\begin{equation}
\int_{s_i}^{s_{th}} \frac{ds}{s}\; s\; f^2_{n(s)}\; \frac{dn}{ds}=
f^2_1 \int_{s_i}^{s_{th}} ds\; \frac{d\ln{n}}{ds} = f^2_1\; \ln{n_{th}}\;.
\label{2.11}
\end{equation}
{}From the other hand, in the leading approximation, one gets
\begin{equation}
\int_{s_i}^{s_{th}} ds\; \frac{\alpha_S}{2\pi}\;
\biggl(\frac{4\mu}{M}\biggr)^2 =
\frac{\alpha_S}{2\pi}\; \biggl(\frac{4\mu}{M}\biggr)^2\; (s_{th}-s_i)\;,
\label{2.12}
\end{equation}
and, further,
\begin{equation}
s_{th}-s_i \simeq 2M \Delta E\;, \label{2.13}
\end{equation}
where $\Delta E = E_{th} - m_Q - m_{Q'}$ is the difference between the
threshold energies for the decay $(Q\bar Q') \to (Q\bar q) + (\bar Q' q)$ and
the $(Q\bar Q')$ pair production.

In HQET \cite{10} one has
\begin{equation}
\Delta E = 2\bar \Lambda + O(1/m_Q)\;, \label{2.14}
\end{equation}
i.e. in the leading approximation one can take $\Delta E \simeq 2\bar \Lambda$,
being a constant value, independent of the heavy quark flavours.

Then one finds
\begin{equation}
\frac{f_1^2}{M} = \frac{\alpha_S}{\pi}
\; \biggl(\frac{4\mu}{M}\biggr)^2 \; \frac{2\bar \Lambda}{\ln{n_{th}}}\;.
\label{2.15}
\end{equation}
Comparing eq.(\ref{38}) and eq.(\ref{2.15}), one can easily find that
in the leading approximation
\begin{equation}
\frac{dM_n}{dn}(n=1) = \frac{2\bar \Lambda}{\ln{n_{th}}}\;. \label{2.16}
\end{equation}
Having derived eq.(\ref{2.16}), one has assumed, that

1) the binding energy of quarks in the $1S$-state is negligibly small, than
the excitation energy of $nS$-levels
\begin{equation}
E_1 \ll \bar \Lambda \sim \frac{dM}{dn}\;, \label{2.17}
\end{equation}

2) the excitation energy of levels is small in comparison with the quark masses
\begin{equation}
\bar \Lambda \sim \frac{dM}{dn} \ll m_Q\;, \label{2.18}
\end{equation}
so that $\sqrt{s} \sim M$,

3) in the leading approximation the hadronic continuum threshold is
determined by the messes of heavy mesons
\begin{equation}
\sqrt{s_{th}} \simeq M_{(Q\bar q)} + M_{(\bar Q' q)} \simeq
m_Q +m_{Q'} + 2 \bar \Lambda \;, \label{2.19a}
\end{equation}

4) the number of states below the threshold is finite and weakly depends on
the heavy quark flavours.

Thus, from eqs.(\ref{2.8}) and (\ref{2.16}) one can conclude that in the
framework of the QCD sum rules, one gets the universal regularity for the
density of $S$-wave quarkonium levels, so, in the leading approximation,
the relation does not depend on the heavy quark flavours
\begin{equation}
\frac{dM_n}{dn}(n=1) = const. \label{2.19}
\end{equation}

\section{Analysis of Relation for Level Density}

Relation (\ref{2.16}) is got in the leading approxomation over the
inverse mass of heavy quarks, when one can neglect the spin-dependent
splittings. Therefore, for the $\bar \Lambda$ estimate we will use the values
of $S$-level masses of the quarkonia $(\bar c c)$ and$(\bar b b)$.

One can easily show

\begin{eqnarray}
m(n^3S_1) & = & m(nS) + \frac{1}{4} \Delta m(nS)\;,\label{2.20}\\
m(n^1S_0) & = & m(nS) - \frac{3}{4} \Delta m(nS)\;,\nonumber
\end{eqnarray}
where $\Delta m(nS)$ is proportional to the leptonic constant squared
$f_{nS}^2$ \cite{1,2,3,11a}, so that from the previous Section it follows
\begin{equation}
\Delta m(nS) = \frac{\Delta m(1S)}{n}\;. \label{2.21}
\end{equation}
{}From the experimental data  one has $\Delta m_\psi(1S) =117$ MeV, and,
taking into account eqs.(\ref{2.20}, \ref{2.21}), one gets
\begin{eqnarray}
m_\psi(1S) & = & 3.068\;\;GeV\;,\nonumber\\
m_\psi(2S) & = & 3.670\;\;GeV\;,\nonumber\\
m_\Upsilon(1S) & = & 9.440\;\;GeV\;,\label{2.22}\\
m_\Upsilon(2S) & = & 10.012\;\;GeV\;,\nonumber
\end{eqnarray}
where we have taken into account that
$$
\Delta m_\Upsilon = \Delta m_\psi\; \frac{\alpha_S(\Upsilon)}
{\alpha_S(\psi)}\;,
$$
and
$$
\alpha_S(Q\bar Q') = \frac{4\pi}{9\ln{\frac{2<T>\mu_{Q\bar Q'}}
{\Lambda_{eff}^2}}}\;,
$$
so that $\alpha_S(\Upsilon)/\alpha_S(\psi)\simeq 3/4$ \cite{12}.
{}From eq.(\ref{2.22}) one has
\begin{eqnarray}
(M_2-M_1)|_\psi & \simeq & 0.602\;\;GeV\;,\label{2.23}\\
(M_2-M_1)|_\Upsilon & \simeq & 0.572\;\;GeV\;,\nonumber
\end{eqnarray}
i.e. in average one has
\begin{equation}
<M_2-M_1> \simeq  0.587\pm 0.015\;\;GeV\;.\label{2.24}
\end{equation}

In the $\Upsilon$-family, where the leading approximation over the
inverse heavy quark mass must be the most reliable, one has
\begin{equation}
n_{th} = 4\;.\label{2.25}
\end{equation}
Then
\begin{equation}
\bar \Lambda  = <M_2-M_1> \simeq  0.587\pm 0.015\;\;GeV\;.\label{2.26}
\end{equation}
Estimate (\ref{2.26}) of the important parameter in HQET is in a good
agreement with the estimates, made in the QCD sum rules for the heavy mesons
\cite{11}
\begin{equation}
\bar \Lambda  \simeq  0.57\pm 0.07\;\;GeV\;.\label{2.27}
\end{equation}
However, one has to note that the offered estimate from the sum rules for
the heavy quarkonium has the accuracy, that surpass the accuracy of estimate
(\ref{2.27}) by about one order of magnitude and it is within the limits of
the accuracy $\delta \bar \Lambda \sim 20$ MeV, that can be achieved, because
of the nonperturbative corrections in QCD \cite{11}.

{}From eq.(\ref{2.26}) one finds the estimate
\begin{equation}
\frac{dM}{dn}(n=1) \simeq  0.85\pm 0.02\;\;GeV\;,\label{2.28}
\end{equation}
that is slightly greater than the estimates, made in papers of ref.\cite{13},
where $dM/dn \simeq 0.75$ GeV was determined in the polinomial interpolation
of heavy quarkonium masses.

\section*{Conclusion}

In the framework of the QCD sum rules, the expression for the density
of $S$-wave levels of the heavy quarkonium is derived
$$
\frac{dM_n}{dn}(n=1) = \frac{2\bar \Lambda}{\ln{n_{th}}}\;,
$$
that in the leading approximation does not depend on the heavy quark flavours.
Here, $\bar \Lambda = m_{(Q\bar q)}-m_Q$ and $n_{th}$ is the number of
$nS$-levels below the threshold of $(Q\bar Q') \to (Q\bar q)+(\bar Q' q)$
decay. The analysis of the spectroscopic data on the charmonium and
bottomonium allows one to do the estimate
$$
\bar \Lambda \simeq  0.587\pm 0.015\;\;GeV\;,
$$
that is in a good agreement with the recent estimates from the QCD sum
rules for the heavy mesons, but the former has the better accuracy.

The derived relation for the heavy quarkonium level density allows one
to do a complete consideration of heavy quarkonium in the specific scheme of
QCD sum rules with no use of the phenomenological data from the
potential models. Such consideration essentially improves the accuracy of
QCD sum rule predictions.

\end{document}